\documentclass[twoside,fleqn]{article}

\usepackage{espcrc2}

\input{psfig}

\newcommand{\beq}{\begin{equation}}
\newcommand{\eeq}{\end{equation}}
\newcommand{\beqn}{\begin{eqnarray}}
\newcommand{\eeqn}{\end{eqnarray}}
\newcommand{\beqns}{\begin{eqnarray*}}
\newcommand{\eeqns}{\end{eqnarray*}}

\newcommand{\hm}{\hspace{-0.05cm}}

\newcommand{\intl}{\int\limits}
\newcommand{\ointl}{\oint\limits}

\newcommand{\e}{\epsilon}
\def\NP{{\it Nucl. Phys.}}
\def\PL{{\it Phys. Lett.}}
\def\PR{{\it Phys. Rev.}}
\def\PRep{{\it Phys. Rep.}}
\def\PRL{{\it Phys. Rev. Lett.}}

\def\ZP{{\it Z. Phys.}}
\def\EPJ{{\it Europ. Phys. J.}}

\def\ea{{\it et al.}}
\def\Cl{Collaboration}
\def\MeVE{~MeV}
\def\pc{$\%$}

\def\sf{spectral function}
\def\sfs{spectral functions}

\def\eee{$e^+e^-$}

\def\eetohad{$e^+e^-\!\rightarrow{\rm hadrons}$}

\def\aqed{$\alpha(s)$}
\def\aqedZ{$\alpha(M_{\rm Z}^2)$}

\def\daqedZ{$\Delta\alpha(M_{\rm Z}^2)$}
\def\daqedh{$\Delta\alpha_{\rm had}(s)$}
\def\daqedhZ{$\Delta\alpha_{\rm had}(M_{\rm Z}^2)$}
\def\amuhad{$a_\mu^{\rm had}$}
\def\aehad{$a_e^{\rm had}$}
\def\MSbar{$\overline{\rm MS}$}
\def\FOPTCI{$\rm FOPT_{\rm CI}$}
\def\ie{{\it i.e.}} 
\def\eg{{\it e.g.}} 
\def\via{via} 

\def\ICHEP{Talk given at the International Conference on High Energy Physics}

\def\be{\begin{equation}}
\def\ee{\end{equation}}
\def\bea{\begin{eqnarray}}
\def\eea{\end{eqnarray}}

\input psfig

\title{Evaluation of \aqedZ\ and $(g-2)_\mu$}

\author{Michel Davier
\address{Laboratoire de l'Acc\'el\'erateur Lin\'eaire,\\
         IN2P3-CNRS et Universit\'e de Paris-Sud, F-91405 Orsay, France \\
         E-mail: davier@lal.in2p3.fr}   
}
\begin{document}

\begin{abstract}
This talk summarizes the recent 
developments in the evaluation of the leading order hadronic contributions 
to the running of the QED f\/ine structure constant \aqed, at $s=M_{\rm Z}^2$,
and to the anomalous magnetic moment of the muon $(g-2)_\mu$. The 
accuracy of the theoretical prediction of these observables
is limited by the uncertainties on the hadronic contributions. 
Signif\/icant improvement has been achieved in a series of new analyses
which is presented historically in three steps: (I), use of $\tau$
spectral functions in addition to \eee\ cross sections, (II), extended
use of perturbative QCD and (III), application of QCD sum rule techniques.
The most precise values obtained are: \daqedhZ\,$=(276.3\pm1.6)\times10^{-4}$,
yielding $\alpha^{-1}(M_{\rm Z}^2)=128.933\pm0.021$, and
$a_\mu^{\rm had}=(692.4\pm6.2)\times 10^{-10}$ with which one
f\/inds for the complete Standard Model prediction
$a_\mu^{\rm SM}=(11\,659\,159.6\pm6.7)\times10^{-10}$.
For the electron $(g-2)_e$, the hadronic contribution is
$a_e^{\rm had}=(187.5\pm1.8)\times 10^{-14}$.
\end{abstract}
\maketitle

\section{ INTRODUCTION }

The running of the QED f\/ine structure constant $\alpha(s)$
and the anomalous magnetic moment of the muon are observables for
which the theoretical precision is limited by second order loop 
ef\/fects from hadronic vacuum polarization. Both quantities are related \via\
dispersion relations to the hadronic production rate in \eee\
annihilation,
\beq
\label{eq_rsigma}
      R(s) = \frac{\sigma_{\rm tot}(e^+e^-\!\rightarrow {\rm hadrons})}
                  {\sigma_0(e^+e^-\!\rightarrow \mu^+\mu^-)}~,
\eeq
with $\sigma_0(e^+e^-\!\rightarrow \mu^+\mu^-) = 4\pi\alpha^2/(3s)$.
While far from quark thresholds and at suf\/f\/iciently high energy
$\sqrt{s}$, $R(s)$ can be predicted by perturbative QCD, theory  
fails when resonances occur, \ie, local quark-hadron duality is broken. 
Fortunately, one can circumvent this drawback by using \eee\ 
annihilation data for $R(s)$ and, as proposed in 
Ref.~\cite{g_2pap}, hadronic $\tau$ decays benef\/itting from the 
largely conserved vector current (CVC), to replace theory in the 
critical energy regions.

There is a strong interest in the electroweak phenomenology to
reduce the uncertainty on \aqedZ\ which used to be a serious limit
to progress in the determination of the Higgs mass from 
radiative corrections in the Standard Model. Table~\ref{sin2} gives
the uncertainties of the experimental and theoretical input expressed
as errors on ${\rm sin}^2\theta_{\rm W}$. Using the
former value~\cite{eidelman} for \aqedZ, the dominant uncertainties
stem from the experimental ${\rm sin}^2\theta_{\rm W}$ determination
and from the running f\/ine structure constant. Thus, any useful 
experimental amelioration on ${\rm sin}^2\theta_{\rm W}$ requires a
better precision of \aqedZ, \ie, an improved determination of
its hadronic contribution.

The anomalous magnetic moment $a_\mu=(g-2)/2$ of the muon is experimentally
and theoretically known to very high accuracy. In addition, the contribution 
of heavier objects to $a_\mu$ relative to the anomalous moment of the electron 
scales as $(m_\mu/m_e)^2\sim4\times10^{4}$. These properties allow an 
extremely sensitive test of the validity of the electroweak theory. 
The present value from the combined $\mu^+$ and $\mu^-$ 
measurements~\cite{bailey},
\beq
    a_\mu \:=\: (11\,659\,230 \pm 85)\times10^{-10}~,
\eeq
is expected to be improved to a precision of at least
$4\times10^{-10}$ by the E821 experiment at 
Brookhaven~\cite{bnl,bnlnew}). Again, the
precision of the theoretical prediction of $a_\mu$ is limited
by the contribution from hadronic vacuum polarization determined
analogously to \aqedZ\ by evaluating a dispersion integral using
\eee\ cross sections and perturbative QCD.
\begin{table}[t]
\caption[.]{ \label{sin2} 
	      Uncertainties of the electroweak input
              expressed in terms of $\Delta{\rm sin}^2\theta_{\rm W}$
              ($\times 10^{-4}$).
              Downward arrows indicate future experimental improvement. }
\setlength{\tabcolsep}{0.1pc}
\vspace{0.2cm}
\begin{center}
\small
\begin{tabular}{lcc} \\ \hline
 Input               & $\Delta{\rm sin}^2\theta_{\rm W}$   
	                    & Uncertainty/Source \\ \hline
 Exp.          &  18  & (LEP+SLD)~\cite{sinth} $\downarrow$ \\
 $\alpha(M_{\rm Z})$ &  23  & $\Delta\alpha^{-1}(M_{\rm Z}^2) = 0.09$~\cite{eidelman} \\
 $m_t$               &  15  & $\Delta m_t = 5.0$~GeV 
                             (CDF+D0)~\cite{topmass} $\downarrow$ \\
 Theory              & 5-10 & 2-loop EW prediction~\cite{degrassi} \\
 $M_H$     & 150  & 65 -- 1000 GeV \\ \hline
\end{tabular}
\end{center}
\end{table}

%
%

\section{RUNNING OF THE QED FINE STRUCTURE CONSTANT}

The running of the electromagnetic f\/ine structure constant \aqed\
is governed by the renormalized vacuum polarization function,
$\Pi_\gamma(s)$. For the spin 1 photon, $\Pi_\gamma(s)$ is given 
by the Fourier transform of the time-ordered product of the 
electromagnetic currents $j_{\rm em}^\mu(s)$ in the vacuum 
$(q^\mu q^\nu-q^2g^{\mu\nu})\,\Pi_\gamma(q^2)=
i\int d^4x\,e^{iqx}\langle 0|T(j_{\rm em}^\mu(x)j_{\rm em}^\nu(0))|0\rangle$.  
With $\Delta\alpha(s)=-4\pi\alpha\,{\rm Re}
\left[\Pi_\gamma(s)-\Pi_\gamma(0)\right]$ and 
$\Delta\alpha(s)=\Delta\alpha_{\rm lep}(s)+\Delta\alpha_{\rm had}(s)$,
which subdivides the running contributions into a leptonic and a 
hadronic part, one has
\beq
    \alpha(s) \:=\: \frac{\alpha(0)}
                         {1 - \Delta\alpha_{\rm lep}(s)
                            - \Delta\alpha_{\rm had}(s)}~,
\eeq
where $4\pi\alpha(0)$ is the square of the electron charge in the 
long-wavelength Thomson limit. 

For the case of interest, $s=M_{\rm Z}^2$, the leptonic contribution 
at three-loop order has been calculated to be~\cite{steinh}
\beq
   \Delta\alpha_{\rm lep}(M_{\rm Z}^2) = 314.97686\times10^{-4}~.
\eeq
Using analyticity and unitarity, the dispersion integral for the 
contribution from hadronic vacuum polarization 
reads~\cite{cabibbo}
\beq\label{eq_int_alpha}
    \Delta\alpha_{\rm had}(M_{\rm Z}^2) \:=\:
        -\frac{\alpha(0) M_{\rm Z}^2}{3\pi}\,
         {\rm Re}\!\!\!\intl_{4m_\pi^2}^{\infty}\!\!ds\,
            \frac{R(s)}{s(s-M_{\rm Z}^2)-i\epsilon}~,
\eeq
and, employing the identity $1/(x^\prime-x-i\epsilon)_{\epsilon\rightarrow0}
={\rm P}\{1/(x^\prime-x)\}+i\pi\delta(x^\prime-x)$, the above
integral is evaluated using the principle value integration 
technique.

%
%
\section{MUON MAGNETIC ANOMALY}

It is convenient to separate the Standard Model prediction for the
anomalous magnetic moment of the muon,
$a_\mu\equiv(g-2)_\mu/2$, into its dif\/ferent contributions,
\beq
    a_\mu^{\rm SM} \:=\: a_\mu^{\rm QED} + a_\mu^{\rm had} +
                             a_\mu^{\rm weak}~,
\eeq
where $a_\mu^{\rm QED}=(11\,658\,470.6\,\pm\,0.2)\times10^{-10}$ is 
the pure electromagnetic contribution (see Ref.~\cite{krause1} and references 
therein), \amuhad\ is the contribution from hadronic vacuum polarization,
and $a_\mu^{\rm weak}=(15.1\,\pm\,0.4)\times10^{-10}
$ accounts for corrections due to exchange of the weak interacting 
bosons up to two loops~\cite{krause1,peris,weinberg}.

Equivalently to \daqedhZ, by virtue of the analyticity of the 
vacuum polarization correlator, the contribution of the hadronic 
vacuum polarization to $a_\mu$ can be calculated \via\ the dispersion 
integral~\cite{rafael}
\beq\label{eq_int_amu}
    a_\mu^{\rm had} \:=\: 
           \frac{\alpha^2(0)}{3\pi^2}
           \intl_{4m_\pi^2}^\infty\!\!ds\,\frac{K(s)}{s}R(s)~,
\eeq
where $K(s)$ denotes the QED kernel~\cite{rafael2}~,
\beqn
    \lefteqn{K(s) \:=\: x^2\left(1-\frac{x^2}{2}\right) \,+\,
                 (1+x)^2\left(1+\frac{1}{x^2}\right)} \nonumber\\
    & &\hspace{0.5cm}\times\left({\rm ln}(1+x)-x+\frac{x^2}{2}\right) \,+\,
                 \frac{(1+x)}{(1-x)}x^2\,{\rm ln}x~,
\eeqn
with $x=(1-\beta_\mu)/(1+\beta_\mu)$ and $\beta_\mu=(1-4m_\mu^2/s)^{1/2}$.
The function $K(s)$ decreases monotonically with increasing $s$. It gives
a strong weight to the low energy part of the integral~(\ref{eq_int_amu}).
About 92\pc\ of the total contribution to \amuhad\ is accumulated at c.m. 
energies $\sqrt{s}$ below 1.8~GeV and 72\pc\ of \amuhad\ is covered by 
the two-pion f\/inal state which is dominated by the $\rho(770)$ 
resonance. Data from vector hadronic $\tau$ decays published by the 
ALEPH Collaboration~\cite{aleph_vsf} provide a precise spectrum for
the two-pion f\/inal state as well as new input for the lesser known 
four-pion f\/inal states. This new information improves signif\/icantly
the precision of the \amuhad\ determination~\cite{g_2pap}.

\section{IMPROVEMENT IN THREE STEPS}
\label{sec_improve}

A very detailed and rigorous evaluation of both \aqedZ\ and $(g-2)_\mu$
was performed by S.~Eidelman and F.~Jegerlehner in 1995~\cite{eidelman}
which since then is frequently used as standard reference. In their
numerical calculation of the integrals~(\ref{eq_int_alpha}) and 
(\ref{eq_int_amu}), the authors use exclusive \eetohad\ cross section 
measurements below 2~GeV c.m. energy, inclusive $R$ measurements 
up to 40~GeV and f\/inally perturbative QCD above 40~GeV.
Their results to which I will later refer are
\beqn
\label{eq_eidel}
     \Delta\alpha_{\rm had}(M_{\rm Z}^2)
         &=& (279.7 \pm 6.5)\times10^{-4}~, \nonumber\\
     a_\mu^{\rm had}
      &=& (702.4 \pm 15.3)\times10^{-10}~.
\eeqn

Due to improvements on the electroweak experimental side these theoretical 
evaluations are insuf\/f\/icient for present needs. Fortunately, new 
data and a better understanding of the underlying QCD phenomena led 
to new and signif\/icantly more accurate determinations of the hadronic
contributions to both observables.

\subsection*{(I) Addition of precise $\tau$ data}
\label{sec_i}

Using the conserved vector current (CVC) it was shown in 
Ref.~\cite{g_2pap} that the addition of precise $\tau$
\sfs, in particular of the $\tau^-\rightarrow\pi^-\pi^0\,\nu_\tau$ 
channel, to the \eee\ annihilation cross section measurements
improves the low-energy evaluation of the 
integrals~(\ref{eq_int_alpha}) and (\ref{eq_int_amu}). Hadronic
$\tau$ decays into $\bar{u}d^\prime$ isovector f\/inal states occur 
via exchange of a virtual $W^-$ boson and have therefore contributions 
from vector and axial-vector currents. On the contrary, f\/inal states
produced via photon exchange in \eee\ annihilation are always vector
but have isovector and isoscalar parts. The CVC
relation between the vector two-pion $\tau$ spectral function 
$v_{J=1}(\tau\rightarrow\pi\pi^0\,\nu_\tau)$ and the 
corresponding isovector \eee\ cross section at energy-squared $s$ reads
\beqns
   \sigma^{I=1}(e^+e^-\to \pi^+\pi^-) = 
         \frac{4\pi\alpha^2(0)}{s}v_{J=1}(\tau\to\pi\pi^0\,\nu_\tau)~,
\eeqns
where $v_{J=1}(\tau\rightarrow\pi\pi^0\,\nu_\tau)$ is essentially the
hadronic invariant mass spectrum normalized to the two-pion
branching ratio and corrected by a kinematic factor appropriate to
$\tau$ decays with hadronic spin $J=1$ ~\cite{aleph_vsf}. The two-pion
cross sections (incl. the $\tau$ contribution) in dif\/ferent 
energy regions are depicted in F\/ig.~\ref{fig_2pi}. Excellent
agreement between $\tau$ and \eee\ data is observed.
For the four pion f\/inal states, isospin rotations must be performed 
to relate the respective $\tau$ final states to the corresponding \eee\ 
topologies~\cite{aleph_vsf}. 

\subsubsection*{\it Effects from $SU(2)$ violation}

Hadronic \sfs\ from $\tau$ decays are directly related to the 
isovector vacuum polarization currents when isospin invariance (CVC) and 
unitarity hold. For this purpose one has to worry whether the breakdown 
of CVC due to quark mass ef\/fects ($m_u \neq m_d$ generating 
$\partial_{\mu}J^\mu\sim(m_u - m_d)$ for a charge-changing hadronic 
current $J^\mu$ between $u$ and $d$ quarks) or unknown isospin-violating 
electromagnetic decays have non-negligible contributions within the 
present accuracy. Expected deviations from CVC due to so-called {\it second 
class currents} as, \eg, the decay $\tau^-\rightarrow\pi^-\eta\,\nu_\tau$ 
where the corresponding \eee\ f\/inal state
$\pi^0\eta$ (C=+1) is strictly forbidden, have estimated branching 
fractions of the order~\cite{etapi} of $(m_u-m_d)^2/m_\tau^2\simeq10^{-5}$, 
while the experimental upper limit amounts to 
B($\tau\rightarrow\pi^-\eta\,\nu_\tau$)\,$\,<1.4\times10^{-4}$~\cite{pdg} at 
95\pc\ CL. $SU(2)$ symmetry breaking caused by electromagnetic interactions
can occur in the $\rho^\pm$--$\rho^0$ masses and widths. Hadronic 
contributions to the $\rho^\pm$--$\rho^0$ width dif\/ference are expected 
to be much smaller since they are proportional to $(m_u-m_d)^2$. 
The total expected $SU(2)$ violation in the $\rho$
width is estimated in Ref.~\cite{g_2pap} to be
$(\Gamma_{\rho^\pm}-\Gamma_{\rho^0})/\Gamma_{\rho}=(2.8 \pm 3.9)\times 10^{-3}$,
yielding the corrections
\beqn
\label{eq_adif}
   \delta a_\mu^{\rm had} 
       &=& 
         -(1.3 \pm 2.0)\times10^{-10}~, \nonumber\\
   \delta \Delta\alpha_{\rm had}^{(5)}(M_{\rm Z}^2)
       &=&
         -(0.09 \pm 0.12)\times10^{-4}~,
\eeqn
to the respective dispersion integrals when using the 
$\tau^-\!\rightarrow\pi^-\pi^0\nu_\tau$ \sf\ in addition to \eee\ data.

\subsubsection*{\it Evaluation of the dispersion integrals~(\ref{eq_int_alpha}) and
                (\ref{eq_int_amu})}

\begin{figure}[h]
\onecolumn
\begin{center}
\hspace{-0.2cm}\psfig{figure=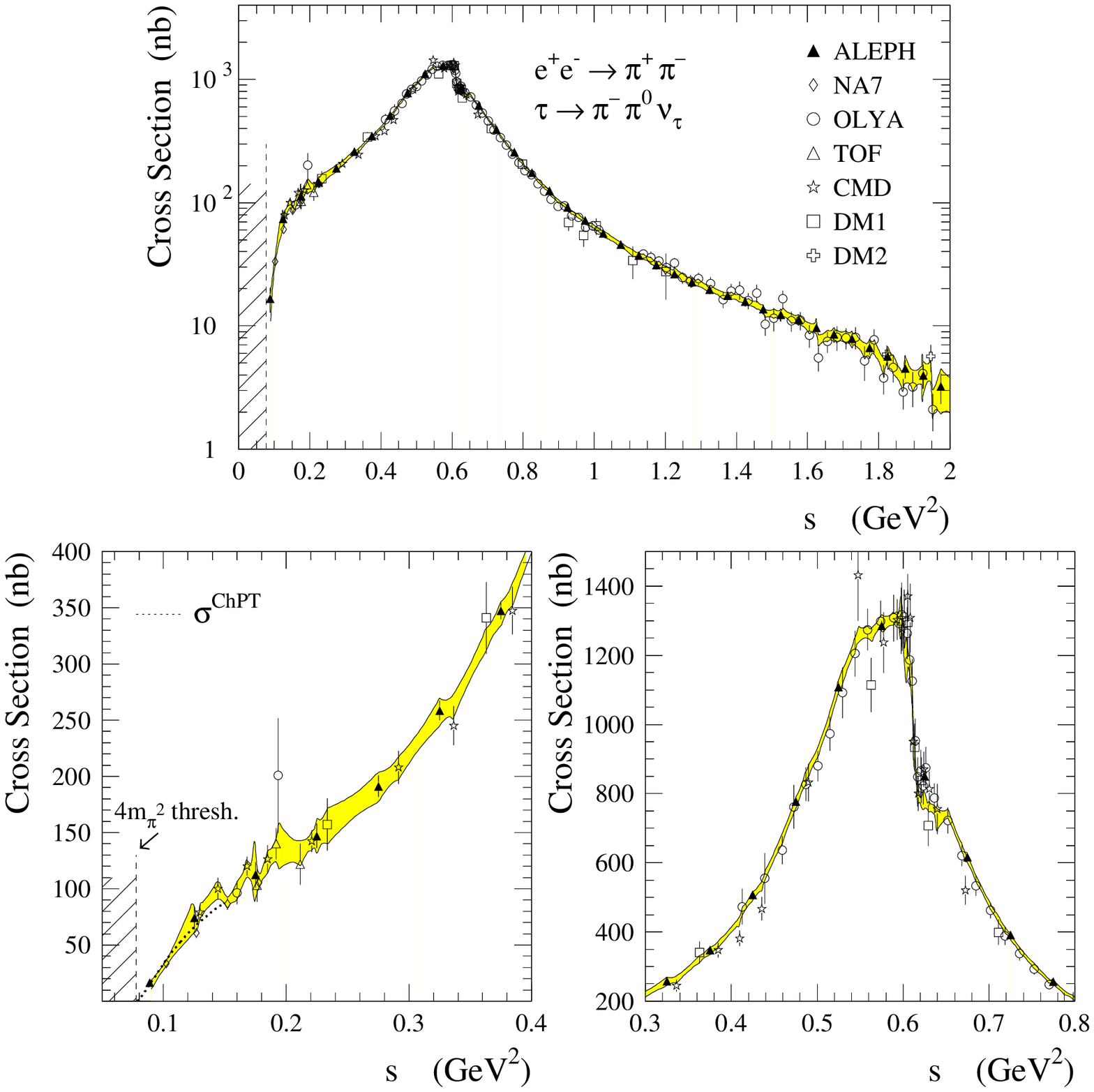,height=3.7in}
\end{center}
\vspace*{-1cm}
\caption[.]{ \label{fig_2pi}
	     Two-pion cross section as a function of the c.m.
             energy-squared. The band represents the result of the averaging 
             procedure described in the text within the diagonal errors. 
             The lower left hand plot shows the chiral expansion of the 
             two-pion cross section used (see Ref.~\cite{g_2pap}).}
\twocolumn
\end{figure}

Details about the non-trivial task of evaluating in a coherent way 
numerical integrals over data points which have statistical and 
correlated systematic errors between measurements {\it and}
experiments are given in Ref.~\cite{g_2pap}. The procedure is based
on an analytical $\chi^2$ minimization, taking into account all initial
correlations, and it provides the averages and the covariances of the 
cross sections from dif\/ferent experiments contributing to a 
certain f\/inal state in a given range of c.m. energies. One then 
applies the trapezoidal rule for the numerical integration of the 
dispersion integrals~(\ref{eq_int_alpha}) and (\ref{eq_int_amu}),
\ie, the integration range is subdivided into
suf\/f\/iciently small energy steps and for each of these steps 
the corresponding covariances (where additional correlations induced by 
the trapezoidal rule have to be taken into account) are calculated. This 
procedure yields error envelopes between adjacent measurements as 
depicted by the shaded bands in Fig.~\ref{fig_2pi}.

\subsubsection*{\it Results}

With the inclusion of the $\tau$ vector \sfs, the hadronic contributions 
to \aqedZ\ and to $a_\mu$ are found to be~\cite{g_2pap}

\beqn
\label{eq_res1}
     \Delta\alpha_{\rm had}(M_{\rm Z}^2)
             & = &  (281.0 \pm 6.2)\times 10^{-4}~,\nonumber \\
     a_\mu^{\rm had} & = & (701.1\,\pm\, 9.4)\times 10^{-10}~, 
\eeqn
with an improvement for $a_\mu^{\rm had}$ of about $40\%$ compared to the
previous evaluation~(\ref{eq_eidel}), while there is only a marginal
improvement of \daqedZ\ for which the dominant uncertainties stem from 
higher energies.

\vspace*{14cm}

\subsection*{(II) Extended theoretical approach}
\label{sec_ii}

The above analysis shows that in order to improve the precision on 
\daqedZ, a more accurate determination of the hadronic cross section between 
2~GeV and 10~GeV is needed. On the experimental side there are
ongoing $R$ measurements performed by the BES Collaboration~\cite{ichep_bes}.
On the other hand, QCD analyses using hadronic $\tau$ decays performed by 
ALEPH~\cite{aleph_asf} and recently by OPAL~\cite{opal_asf} 
revealed excellent applicability of the Wilson {\it Operator Product 
Expansion} (OPE)~\cite{wilson} (also called SVZ 
approach~{\cite{svz}), organizing perturbative and nonperturbative contributions
to a physical observable using the concept of global quark-hadron duality, at
the scale of the $\tau$ mass, $M_\tau\simeq1.8~{\rm GeV}$. Using moments of 
spectral functions, dimensional nonperturbative operators contributing  to the 
$\tau$ hadronic width have been f\/itted simultaneously and turned out to be
small. This encouraged the authors of Ref.~\cite{alphapap} to apply 
a similar approach based on spectral moments to determine the size of
the nonperturbative contributions to integrals over total cross sections in 
\eee\ annihilation, and to f\/igure out whether or not the OPE, \ie, global 
duality is a valid approach at relatively low energies.

\subsubsection*{\it Theoretical prediction of $R(s)$}

The optical theorem relates the total hadronic cross section in \eee\ 
annihilation, $R(s_0)$, at a given energy-squared, $s_0$, to the absorptive 
part of the photon vacuum polarization correlator
\beq
\label{eq_rimpi}
     R(s_0) = 12\pi{\rm Im}\Pi(s_0+i\epsilon)~.
\eeq
Perturbative QCD predictions up to next-to-next-to leading order
($\alpha_s^3$) as well as second order quark mass corrections far from
the production threshold and
the f\/irst order dimension $D=4$ nonperturbative term are available 
for the Adler $D$-function~\cite{adler},
which is the logarithmic derivative of the correlator $\Pi$,
carrying all physical information:
\beq
\label{eq_adler}
     D(s) = - 12 \pi^2 s\frac{d\Pi(s)}{ds}~.
\eeq
This yields the relation
\beq
\label{eq_radler}
     R(s_0) = \frac{1}{2\pi i}
              \hm\ointl_{|s|=s_0}\hm\hm\frac{ds}{s} D(s)~,
\eeq
where the contour integral runs counter-clockwise around the 
circle from $s=s_0+i\e$ to $s=s_0-i\e$.
The Adler function is given by~\cite{3loop,kataev,bnp}
\beqn
\label{eq_d}
   \lefteqn{D_{f_i}(-s) = N_C\sum_f Q_f^2
            \Bigg\{}\nonumber \\
         & & \hspace{0.0cm}
            1 + d_0\,a_s
                     + d_1\,a_s^2
                     + \tilde{d}_2\,a_s^3 
                     \nonumber \\
         & & \hspace{0.0cm} 
           -\; \frac{m_f^2(s)}{s}
                \left( 6 + 28\,a_s
                        + (295.1-12.3\,n_f)\,a_s^2       
                 \right)  \nonumber \\
         & & \hspace{0.0cm} 
           +\; \frac{2\pi^2}{3}\left(1 - \frac{11}{18}\,a_s
                         \right)\frac{\left\langle\frac{\alpha_s}{\pi} 
                                           GG\right\rangle}{s^2} 
      \nonumber \\
         & & \hspace{0.0cm} 
           + \;8\pi^2\left(1 - a_s
                  \right)\frac{\langle m_f\bar{q_f}q_f\rangle}{s^2}
      \nonumber \\
         & & \hspace{0.0cm} 
           + \;\frac{32\pi^2}{27}a_s
              \sum_k\frac{\langle m_k\bar{q_k}q_k\rangle}{s^2}
      \nonumber \\
         & & \hspace{0.0cm} 
           + \;12\pi^2\frac{\langle{\cal O}_6\rangle}{s^3}
            \;+\; 16\pi^2\frac{\langle{\cal O}_8\rangle}{s^4}
      \Bigg\}~, 
\eeqn
where additional logarithms occur when $\mu^2\neq s$, $\mu$ 
being the renormalization scale\footnote 
{
   The negative energy-squared in $D(-s)$ of Eq.~(\ref{eq_d})
   is introduced when continuing the Adler function from the spacelike
   Euclidean space, where it is originally def\/ined, to the timelike
   Minkowski space by virtue of its analyticity property.
}
and $a_s=\frac{\alpha_s(s)}{\pi}$.
The coef\/f\/icients of the massless perturbative part are 
$d_0=1$, $d_1=1.9857 - 0.1153\,n_f$,
$\tilde{d}_2=d_2+\beta_0^2\pi^2/48$ with $\beta_0=11-2n_f/3$,
$d_2=-6.6368-1.2001\,n_f-0.0052\,n_f^2-1.2395\,(\sum_f Q_f)^2/N_C\sum_f Q_f^2$
and $n_f$ being the number of involved quark f\/lavours.
The nonperturbative operators in Eq.~(\ref{eq_d}) are the gluon condensate, 
$\langle(\alpha_s/\pi) GG\rangle$, and the quark condensates,
$\langle m_f\bar{q_f}q_f\rangle$. The latter obey approximately the PCAC 
relations
\beqn
\label{eq_pcac}
     (m_u + m_d)\langle\bar{u}u + \bar{d}d\rangle
       & \simeq & - 2 f_\pi^2 m_\pi^2~, \nonumber \\
     m_s\langle\bar{s}s\rangle 
       & \simeq & - f_\pi^2 (m_K^2-m_\pi^2)~,
\eeqn
with the pion decay constant $f_\pi=(92.4\pm0.26)$\MeVE~\cite{pdg}.
In the chiral limit the equations $f_\pi=f_K$ and
$\langle\bar{u}u\rangle=\langle\bar{d}d\rangle=\langle\bar{s}s\rangle$
hold. The complete dimension $D=6$ and $D=8$ operators are parametrized 
phenomenologically in Eq.~(\ref{eq_d}) using the saturated vacuum 
expectation values $\langle{\cal O}_6\rangle$ and 
$\langle{\cal O}_8\rangle$, respectively.

Although the theoretical prediction of $R$ using Eqs.~(\ref{eq_radler})
and (\ref{eq_d}) assumes {\it local} duality and therefore suf\/fers from 
unpredicted low-energy resonance oscillations, the following 
integration, Eqs.~(\ref{eq_int_alpha})/(\ref{eq_int_amu}), turns 
the hypothesis into {\it global} duality, \ie, 
the nonperturbative oscillations are averaged over the energy 
spectrum. However, a systematic uncertainty is introduced through the cut
at explicitly 1.8~GeV so that non-vanishing oscillations could give rise to 
a bias after integration. The associated systematic error is estimated
in Ref.~\cite{apap} by means of f\/itting dif\/ferent oscillating curves 
to the data around the cut region, yielding the error estimates
$\Delta(\Delta \alpha_{\rm had}(M_{\rm Z}^2))=0.15\times10^{-4}$ and
$\Delta a_\mu^{\rm had} = 0.24\times10^{-10}$,
from the comparison of the integral over the oscillating simulated
data to the OPE prediction. These numbers are added as systematic
uncertainties to the corresponding low-energy integrals.

In asymptotic energy regions we use the formulae of Ref.~\cite{kuhn1}
which include complete quark mass corrections up to order $\alpha_s^2$ to 
evaluate the perturbative prediction of $R(s)$ entering into the
integrals~(\ref{eq_int_alpha}) and (\ref{eq_int_amu}).

\subsubsection*{\it Theoretical uncertainties}

Details about the parameter errors used to estimate the uncertainties 
accompanying the theoretical analysis are given in Refs.~\cite{alphapap,apap}.
Theoretical uncertainties arise from essentially three sources
\begin{itemize}
  \item[(\it i)] {\it The perturbative prediction.} The estimation
                of theoretical errors of the perturbative series is
                strongly linked to its truncation at f\/inite order
                in $\alpha_s$. This introduces a non-vanishing dependence 
                on the choice of the renormalization scheme and the 
                renormalization scale. Furthermore, one has
                to worry whether the missing four-loop order 
                contribution $d_3\,a_s^4$ gives rise to
                large corrections to the perturbative series.
	        An additional uncertainty stems from the ambiguity
	        between the results on $R$ obtained using contour-improved 
                f\/ixed-order perturbation theory (\FOPTCI) and FOPT only 
                (see Ref.~\cite{aleph_asf}). The value
                $\alpha_s(M_Z^2) = 0.1201 \pm 0.0020$ is taken, as the average
                between the results from $\tau$ hadronic decays and
                the Z hadronic width which have essentially uncorrelated
                uncertainties~\cite{moriond98}.
  \item[(\it ii)] {\it The quark mass correction.} Since a theoretical 
                evaluation of the integrals~(\ref{eq_int_alpha}) and 
                (\ref{eq_int_amu}) is only applied far from quark production 
                thresholds, quark mass corrections and the corresponding 
                errors are small.
  \item[(\it iii)] {\it The nonperturbative contribution}. In order to
                detach the measurement from theoretical constraints
                on the nonperturbative parameters of the OPE, the dominant
                dimension $D=4,6,8$ terms are determined experimentally
                by means of a simultaneous f\/it of weighted integrals over 
                the inclusive low energy \eee\ cross section, so-called spectral 
	        moments, to the theoretical prediction obtained from 
	        Eq.~(\ref{eq_d}). Small uncertainties are introduced from 
                possible deviations from the PCAC relations~(\ref{eq_pcac}).
\end{itemize}

The spectral moment f\/it of the nonperturbative operators results in 
a very small contribution from the OPE power terms to the lowest moment 
at the scale of $1.8~{\rm GeV}$ (repeated and conf\/irmed at $2.1~{\rm GeV}$), 
as expected from the $\tau$ analyses~\cite{aleph_asf,opal_asf}. The value of
$\langle\frac{\alpha_s}{\pi} GG\rangle=(0.037\pm0.019)~{\rm GeV}^4$
found for the gluon condensate is compatible with other 
evaluations~\cite{reinders,bertlmann}. The analysis proves that 
global duality holds at $1.8~{\rm GeV}$ and nonperturbative ef\/fects
contribute only negligibly, so that above this energy perturbative QCD 
can replace the rather imprecise data in the dispersion 
integrals~(\ref{eq_int_alpha}) and (\ref{eq_int_amu}).

\subsubsection*{\it Results}

The $R(s)$ measurements and the corresponding theoretical 
prediction are shown in F\/ig.~\ref{fig_r}. The wide shaded bands indicate 
the regions where data are used instead of theory to evaluate the 
dispersion integrals, namely below 1.8~GeV and at $c\bar{c}$ threshold
energies. Good agreement between data and QCD is found above 8~GeV, 
while at lower energies systematic deviations are observed. The $R$ 
measurements in this region are essentially provided by the 
$\gamma\gamma2$~\cite{E_78} and MARK~I~\cite{E_96} Collaborations. 
MARK~I data above 5~GeV lie systematically above the measurements of 
the Crystal Ball~\cite{CB} and MD1~\cite{MD1} Collaborations as well 
as above the QCD prediction.
\newpage

\begin{figure}[t]
\onecolumn
\begin{center}
\hspace{0.1cm}\psfig{figure=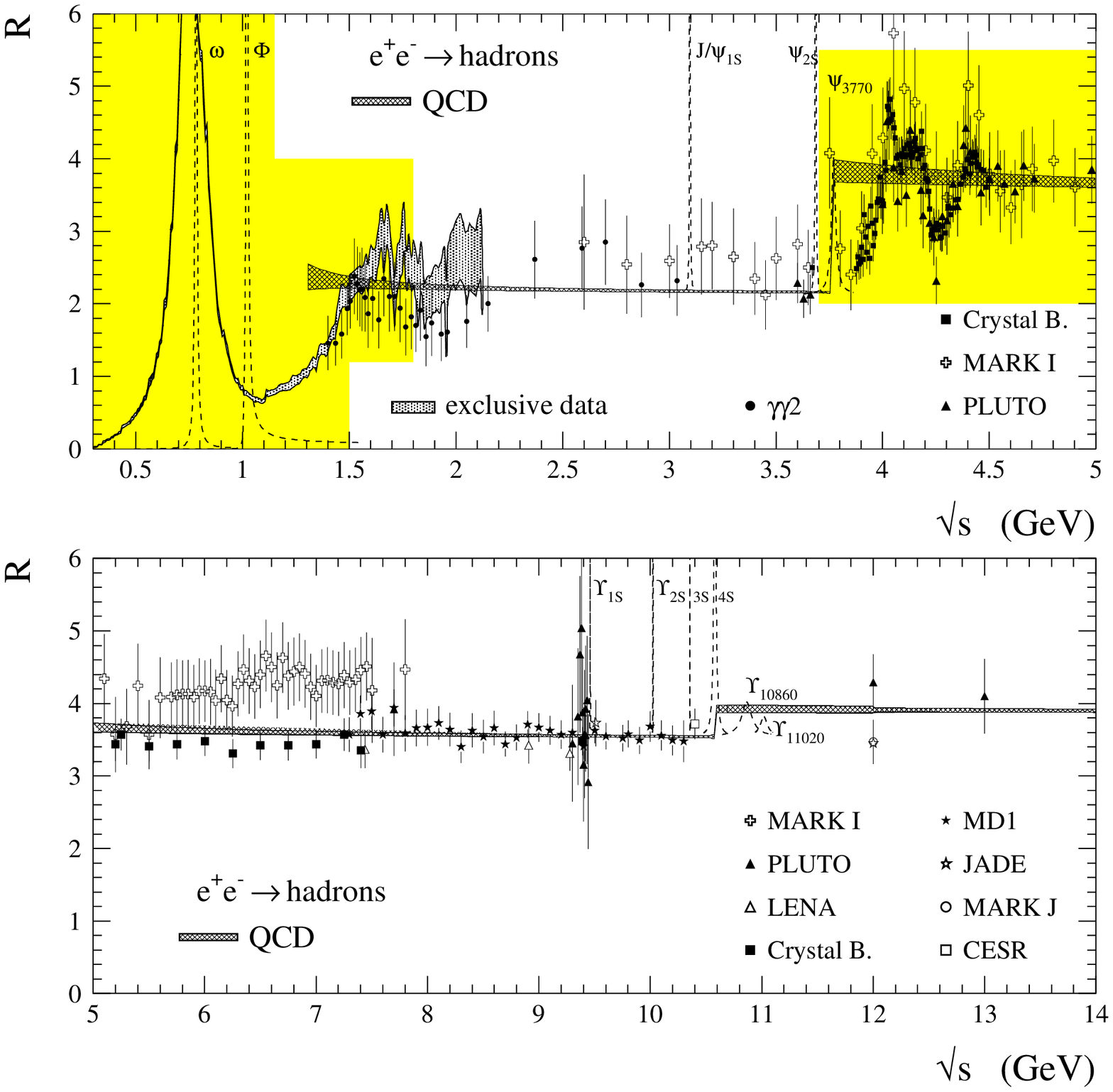,height=3.7in}
\end{center}
\vspace{-1cm}
\caption[.]{\label{fig_r}
            Inclusive hadronic cross section ratio in \eee\
            annihilation versus the c.m. energy $\sqrt{s}$. 
            Additionally shown is the QCD prediction of the continuum 
            contribution as explained in the text. The shaded areas 
            indicate regions were experimental data are used for the 
            evaluation of \daqedhZ\ and \amuhad\ in addition to the 
            measured narrow resonance parameters. The exclusive 
            \eee\ cross section measurements at low c.m. energies
            are taken from DM1,DM2,M2N,M3N,OLYA,CMD,ND and
            $\tau$ data from ALEPH (see Ref.~\cite{g_2pap}
            for detailed information).}
\twocolumn
\vspace{-1.4cm}
\end{figure}

The combination of the theoretical and experimental evaluations of the 
integrals yields the results~\cite{alphapap}
\beqn
\label{eq_res2}
     \Delta\alpha_{\rm had}(M_{\rm Z}^2)
         &=& (277.8 \pm 2.2_{\rm exp} \pm 1.4_{\rm theo})\times10^{-4},~~~~ \nonumber\\
     a_\mu^{\rm had}
      &=& (695.1 \pm 7.5_{\rm exp} \pm 0.7_{\rm theo})\times10^{-10}~,~~~~
\eeqn
with a signif\/icant improvement by more than a factor of two for \daqedZ,
and a $20\%$ better accuracy on $a_\mu^{\rm had}$ compared to the 
numbers~(\ref{eq_res1}).

The authors of Ref.~\cite{kuhnstein} improved the above 
analysis in the charm region by normalizing experimental results in the
theoretically not accessible region (at least locally) so that they match 
perturbative QCD at safe energies below and above the occurrence of 
resonances. 
\pagebreak
\vspace*{12.5cm}

The so-renormalized data show excellent agreement among 
dif\/ferent experiments which supports the hypothesis made that experimental 
systematic errors are completely correlated over the whole involved energy 
regime. The result (after correcting for the small top quark contribution)
reads~\cite{kuhnstein}
\beq
\label{eq_kuhnstein}
     \Delta\alpha_{\rm had}(M_{\rm Z}^2)
         = (276.7 \pm 1.7)\times10^{-4}~.
\eeq

Another, very elegant method based on an analytical calculation of the 
unsubtracted dispersion relation, corresponding to the subtracted
integral~(\ref{eq_int_alpha}), was presented in Ref.~\cite{erler}.
Only the low-energy pole contribution is taken from data,
while the contribution from higher energies is calculated analytically
using the two-point correlation function given in Ref.~\cite{kuhn1},
and the renormalization group equations for the running quantities.
This leads to the precise result~\cite{erler}
\beq
\label{eq_erler}
     \Delta\alpha_{\rm had}(M_{\rm Z}^2)
         = (277.2 \pm 1.9)\times10^{-4}~.
\eeq

Both numbers~(\ref{eq_kuhnstein}) and (\ref{eq_erler}) are in agreement
with \daqedZ\ from Eq.~(\ref{eq_res2}).

\subsection*{(III) Constraints from QCD sum rules}
\label{sec_iii}

It was shown in Ref.~\cite{apap} that the previous determinations 
can be further improved by using f\/inite-energy QCD sum rule techniques 
in order to access theoretically energy regions where locally perturbative 
QCD fails. This idea was f\/irst presented in Ref.~\cite{schilcher}. In 
principle, the method uses no additional assumption beyond those applied in 
Section~\ref{sec_ii}. However, parts of the dispersion integrals evaluated at 
low-energy and the $c\bar{c}$ threshold are obtained from values of the Adler
$D$-function itself, for which local quark-hadron duality is assumed to hold.
One therefore must perform an evaluation at rather high energies (3~GeV for 
$u,d,s$ quarks and 15~GeV for the $c$ quark contribution have been chosen in 
Ref.~\cite{apap}) to suppress deviations from local duality due to 
nonperturbative phenomena.

The idea of the approach is to reduce the data contribution to the
dispersion integrals by subtracting analytical functions from the singular
integration kernels in Eqs.~(\ref{eq_int_alpha}) and (\ref{eq_int_amu}),
and adding the subtracted part subsequently by using theory only.
Two approaches have been applied in Ref.~\cite{apap}: f\/irst, a method
based on spectral moments is def\/ined by the identity
\beqn
\label{eq_f}
     \lefteqn{F = 
          \intl_{4m_\pi^2}^{s_0}\!\!ds\,R(s)\left[f(s) - p_{n}(s)\right]}
          \nonumber \\
     & & \hspace{0.6cm}    
             +\; \frac{1}{2\pi i}\!\!\ointl_{|s|=s_0}\!\!\!\frac{ds}{s}
             \left[P_{n}(s_0) - P_{n}(s)\right] D_{uds}(s)~,
\eeqn
with $P_{n}(s)=\int_0^sdt\,p_{n}(t)$ and 
$f(s)=\alpha(0)^2K(s)/(3\pi^2s)$ for 
$F\equiv a_{\mu,\,[2m_\pi,\;\sqrt{s_0}]}^{\rm had}$,
as well as $f(s)=\alpha(0)M_{\rm Z}^2/(3\pi s(s-M_{\rm Z}^2))$ for 
$F\equiv \Delta\alpha_{\rm had}(M_{\rm Z}^2)_{[2m_\pi,\;\sqrt{s_0}]}$.
The analytic functions $p_{n}(s)$ approximate the kernel $f(s)$ in 
order to reduce the contribution of the f\/irst integral in 
Eq.~(\ref{eq_f}) which has a singularity at $s=0$ and 
is thus evaluated using experimental data. The second integral in 
Eq.~(\ref{eq_f}) can be calculated theoretically in the framework of 
the OPE. The functions $p_{n}(s)$ are 
chosen in order to reduce the {\it uncertainty} of the data integral.
This approximately coincides with a low residual value of the 
integral, \ie, a good approximation of the integration kernel $f(s)$
by the $p_{n}(s)$ def\/ined as~\cite{apap}
\beq
\label{eq_pol}
    p_{n}(s) \equiv \sum_{i=1}^n c_i
                    \left(1 - \left(\frac{s}{s_0}\right)^{\!\!i}\right)~,
\eeq
with the form $(1-s/s_0)$ in order to ensure a vanishing integrand at
the crossing of the positive real axis where the validity of the OPE 
is questioned~\cite{svz}. Polynomials of order $s^n$ involve leading order
nonperturbative contributions of dimension $D=2(n+1)$. The analysis is 
therefore restricted to the linear $n=1$ case only.

A second approach uses the dispersion relation of the Adler $D$-function
\beq
\label{eq_dispd}
    D_f(Q^2) = Q^2\!\!\intl_{4m_f^2}^\infty \!\!ds\,\frac{R_f(s)}{(s+Q^2)^2}~,
\eeq
for space-like $Q^2=-q^2$ and the quark f\/lavour $f$. The 
above integrand approximate the integration kernels in 
Eqs.~(\ref{eq_int_alpha}) and (\ref{eq_int_amu}), so that the 
modif\/ied Eq.~(\ref{eq_f}) reads
\beqn
\label{eq_disp}
  \lefteqn{F = 
        \intl_{4m_\pi^2}^{s_0}\!\!ds\,R^{\rm Data}(s)
               \left[f(s) - \frac{A_FQ^2}{(s+Q^2)^2}\right]} \nonumber \\
    & & \hspace{0.5cm}
         +\; A_F\left(D_{uds}(Q^2) - Q^2\!\intl_{s_0}^\infty\!ds\,
           \frac{R_{uds}^{\rm QCD}(s)}{(s+Q^2)^2}\right)~,
\eeqn
with a normalization constant $A_F$ to be optimized for both \daqedZ\ and 
\amuhad.

In both approaches, a compromise must be obtained between uncertainties of
experimental and theoretical origins. As the subtracted contribution
increases, the experimental error diminishes as expected. However, this
improvement is spoiled by a correspondingly larger theoretical error. The
procedure followed is designed to minimize the total 
uncertainty~\cite{apap}.

\subsubsection*{\it Results}

A $\chi^2$ f\/it taking into account the experimental and theoretical 
correlations between the polynomial moments yields for the f\/irst (spectral
moment) approach (hadronic contribution from 
$2m_\pi$ to 1.8~GeV)~\cite{apap}{\small
\beqns
\label{eq_fitres}
   \Delta\alpha_{\rm had}(M_{\rm Z}^2)_{[0,1.8]}
      &=& (56.53 \pm 0.73_{\rm exp} \pm 0.39_{\rm th})\times 10^{-4}~,\nonumber\\
   a_{\mu,\,[0,1.8]}^{\rm had}
      &=& (634.3 \pm 5.6_{\rm exp} \pm 2.1_{\rm th})\times 10^{-10}~,\nonumber 
\eeqns
}while the dispersion relation approach gives 
($\sqrt{Q^2}=3~{\rm GeV}$)~\cite{apap}{\small
\beqns
   \Delta\alpha_{\rm had}(M_{\rm Z}^2)_{[0,1.8]}
      &=& (56.36 \pm 0.70_{\rm exp} \pm 0.18_{\rm th})\times 10^{-4}~,\nonumber\\
   a_{\mu,\,[0,1.8]}^{\rm had}
      &=& (632.5 \pm 6.2_{\rm exp} \pm 1.6_{\rm th})\times 10^{-10}~.\nonumber
\eeqns
}Only the most precise of the above numbers are used for the f\/inal results.
The above theory-improved results can be compared to the corresponding
pure experimental values,
$\Delta\alpha_{\rm had}(M_{\rm Z}^2)_{[0,1.8]}=(56.77\pm1.06)\times 10^{-4}$
and $a_{\mu,\,[0,1.8]}^{\rm had}=(635.1 \pm 7.4)\times 10^{-10}$,
showing clear improvement.

For the charm threshold region only the approach~(\ref{eq_disp}) is 
used giving for the hadronic contributions from 3.7~GeV to 5~GeV
($\sqrt{Q^2}=15~{\rm GeV}$)~\cite{apap}{\small
\beqn
   \Delta\alpha_{\rm had}(M_{\rm Z}^2)_{[3.7,5]}
      &=& (24.75 \pm 0.84_{\rm exp} \pm 0.50_{\rm th})\times 10^{-4}~,\nonumber\\
   a_{\mu,\,[3.7,5]}^{\rm had}
      &=& (14.31 \pm 0.50_{\rm exp} \pm 0.21_{\rm th})\times 10^{-10}~,
 \nonumber
\eeqn
}for which compared with the pure data results,
$\Delta\alpha_{\rm had}(M_{\rm Z}^2)_{[3.7,5]}=(25.04\pm1.21)\times 10^{-4}$
and $a_{\mu,\,[3.7,5]}^{\rm had}=(14.44 \pm 0.62)\times 10^{-10}$,
only a slight improvement is observed.

\section{FINAL RESULTS}

Table~\ref{tab_alphares} shows the experimental and theoretical 
evaluations of \daqedhZ, \amuhad\ and \aehad\ for the respective 
energy regimes\footnote
{
   The evaluation of \aehad\ follows the same procedure as \amuhad.
}. 
Experimental errors between dif\/ferent lines are
assumed to be uncorrelated, whereas theoretical errors, but
those from $c\bar{c}$ and $b\bar{b}$ thresholds which are quark mass
dominated, are added linearily.

According to Table~\ref{tab_alphares}, the combination of the theoretical
and experimental evaluations of the integrals~(\ref{eq_int_alpha}) 
and (\ref{eq_int_amu}) yields the f\/inal results
\beqn
\label{eq_res3}
   \Delta\alpha_{\rm had}(M_{\rm Z}^2) 
      &=& (276.3 \pm 1.1_{\rm exp} \pm 1.1_{\rm th})\times10^{-4}~, \nonumber\\
   \alpha^{-1}(M_{\rm Z}^2) 
      &=& 128.933 \pm 0.015_{\rm exp} \pm 0.015_{\rm th}~, \\[0.3cm] 
   a_\mu^{\rm had} 
      &=& (692.4 \pm 5.6_{\rm exp} \pm 2.6_{\rm th})\times10^{-10}~, \nonumber\\
   a_\mu^{\rm SM}
      &=& (11\,659\,159.6 \pm 5.6_{\rm exp} \pm 3.7_{\rm th})\times10^{-10}~,
\nonumber
\eeqn

and $a_e^{\rm had}=(187.5\pm1.7_{\rm exp}\pm0.7_{\rm th})\times10^{-14}$ 
for the leading order hadronic contribution to $a_e$. 
The improvement for \aqedZ\ compared to the previous 
results~(\ref{eq_res2}) amounts to $40\%$ and $a_\mu^{\rm had}$ is 
about $17\%$ more precise than~(\ref{eq_res2}).

The total $a_\mu^{\rm SM}$ value includes an additional contribution
from non-leading order hadronic vacuum polarization summarized in 
Refs.~\cite{krause2,g_2pap} to be 
$a_\mu^{\rm had}[(\alpha/\pi)^3]=(-10.0\pm0.6)\times10^{-10}$.
Also the light-by-light scattering (LBLS) contribution has recently been 
reevaluated in Refs.~\cite{kinolight} and \cite{light2} of which the 
average $\langle a_\mu^{\rm had}[{\rm LBLS}]\rangle=(-8.5\pm2.5)\times10^{-10}$
is used here.

Figure~\ref{fig_results_all} shows a 
compilation of published results for the hadronic contributions
to \aqedZ\ and $a_\mu$. Some authors give the contribution 
for the f\/ive light quarks only and add the top quark part separately. 
This has been corrected for in F\/ig.~\ref{fig_results_all}.

\section{THE MASS OF THE STANDARD MODEL HIGGS BOSON}

The new precise result (\ref{eq_res3}) for \aqed\ is exploited 
to repeat the global electroweak f\/it in order to adjust the mass of the
Standard Model Higgs boson, $M_H$. The corresponding uncertainty
on ${\rm sin}^2\theta_{\rm W}$ now reaches $5 \times 10^{-4}$,
well below the experimental accuracy on this quantity and on $m_t$ 
(see Table \ref{sin2}). The prediction of the 
Standard Model is obtained from the ZFITTER electroweak 
library~\cite{leplib} and the experimental input is taken from
the latest review~\cite{sinth}. The standard 
value~\cite{eidelman} of \aqed\ yields
\beqn
   log(M_H) &=& 1.92^{+0.32}_{-0.41}~,\nonumber\\
   M_H      &=& (84^{+91}_{-51})~{\rm GeV/c}^2~, 
\eeqn
while the improved determination (\ref{eq_res3}) provides
 \beqn
   log(M_H) &=& 2.02^{+0.23}_{-0.25}~,\nonumber\\
   M_H      &=& (105^{+73}_{-46})~{\rm GeV/c}^2~. 
\eeqn

\begin{table*}[t]
\caption[.]{\label{tab_alphares}
            Contributions to \daqedhZ, \amuhad\ and to \aehad\ from the 
            dif\/ferent energy regions. The subscripts in the f\/irst column
            give the quark flavours involved in the calculation.}
\setlength{\tabcolsep}{0.4pc}
\vspace{0.2cm}
\begin{center}
{\small
\begin{tabular*}{\textwidth}{lccc} \hline \\[-0.33cm] 
Energy~(GeV)
                         & $\Delta \alpha_{\rm had}(M_{\rm Z}^2)\times10^{4}$
                              & $a_\mu^{\rm had}\times10^{10}$ 
                                   & $a_e^{\rm had}\times10^{14}$ 
\\[0.07cm]
\hline \\[-0.33cm]
$(2m_\pi$ -- $1.8)_{uds}$
                         & $56.36\pm0.70_{\rm exp}\pm0.18_{\rm th}$
                              & $634.3\pm5.6_{\rm exp}\pm2.1_{\rm th}$ 
                                   & $173.67\pm1.7_{\rm exp}\pm0.6_{\rm th}$ 
\\[0.07cm]
$(1.8$ -- $3.700)_{uds}$ & $24.53\pm0.28_{\rm th}$  
                              & $33.87\pm0.46_{\rm th}$     
                                   & $8.13\pm0.11_{\rm th}$
\\[0.07cm]
$\psi(1S,2S,3770)_c$ 
$+~(3.7$ -- $5)_{udsc}$
                         &$24.75\pm0.84_{\rm exp}\pm0.50_{\rm th}$
                              & $14.31\pm0.50_{\rm exp}\pm0.21_{\rm th}$
                                   & $3.41\pm0.12_{\rm exp}\pm0.05_{\rm th}$
\\[0.07cm]
$(5$ -- $9.3)_{udsc}$    & $34.95\pm0.29_{\rm th}$   
                              & $6.87\pm0.11_{\rm th}$ 
                                   & $1.62\pm0.03_{\rm th}$ 
\\[0.07cm]
$(9.3$ -- $12)_{udscb}$
                         &$15.70\pm0.28_{\rm th}$
                              & $1.21\pm0.05_{\rm th}$ 
                                   & $0.28\pm0.02_{\rm th}$ 
\\[0.07cm]
$(12$ -- $\infty)_{udscb}$
                         & $120.68\pm0.25_{\rm th}$   
                              & $1.80\pm0.01_{\rm th}$ 
                                   & $0.42\pm0.01_{\rm th}$ 
\\[0.07cm]
$(2m_t$ -- $\infty)_t$
                         &$-0.69\pm0.06_{\rm th}$  
                              & $\approx0 $ 
                                   & $\approx0 $ 
\\[0.07cm]
\hline\\[-0.33cm]
$(2m_\pi$ -- $\infty)_{udscbt}$
                         & $276.3\pm1.1_{\rm exp}\pm1.1_{\rm th}$
                              & $692.4\pm5.6_{\rm exp}\pm2.6_{\rm th}$ 
                                   & $187.5\pm1.7_{\rm exp}\pm0.7_{\rm th}$ 
\\[0.07cm]
\hline
\end{tabular*}
}
\end{center}
\vspace*{1cm}
\end{table*}

\begin{figure*}
\setlength{\tabcolsep}{-0.4pc}
\begin{tabular*}{\textwidth}{cc}
\psfig{figure=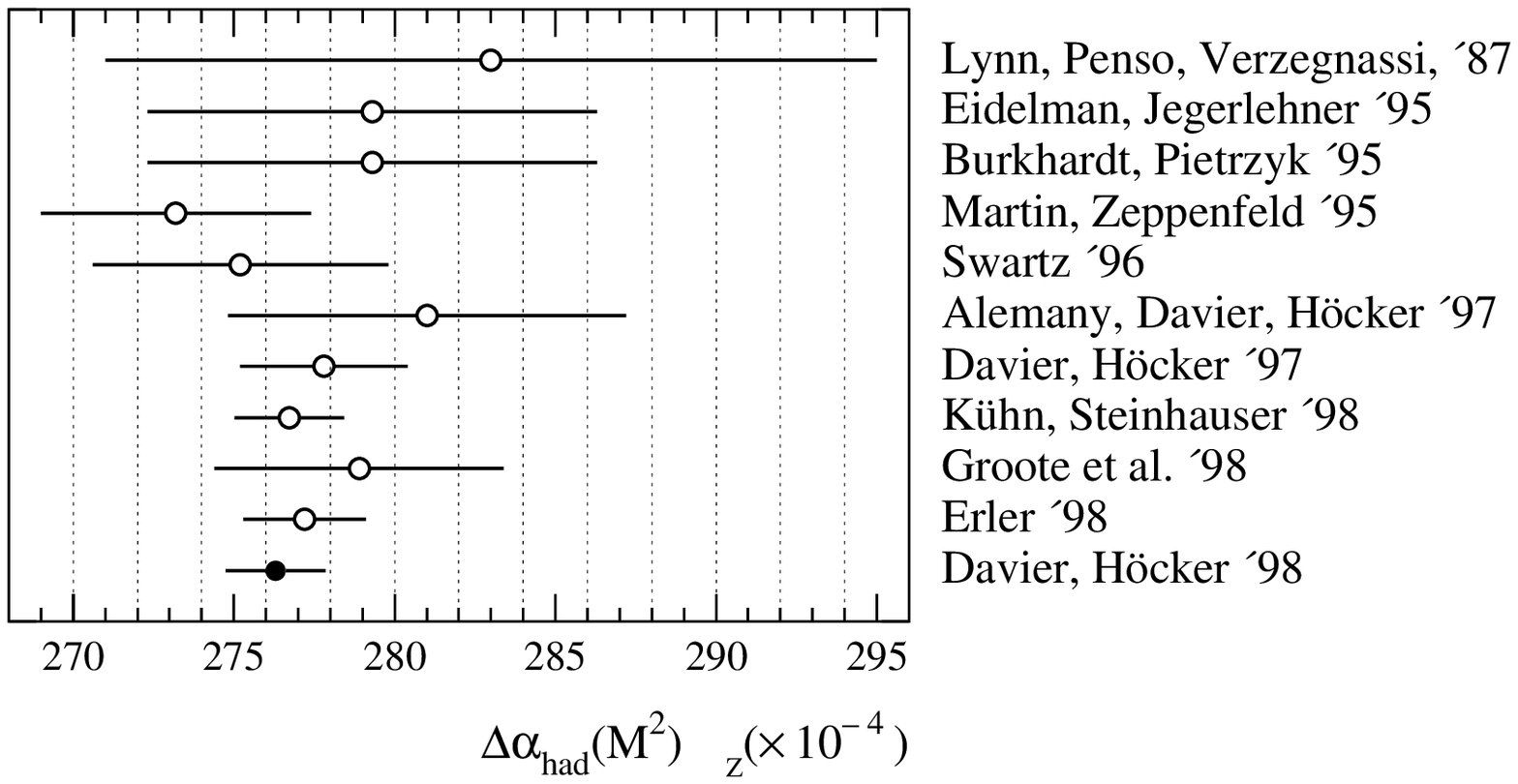,height=1.65in}&
\hspace*{-0.5cm}
\psfig{figure=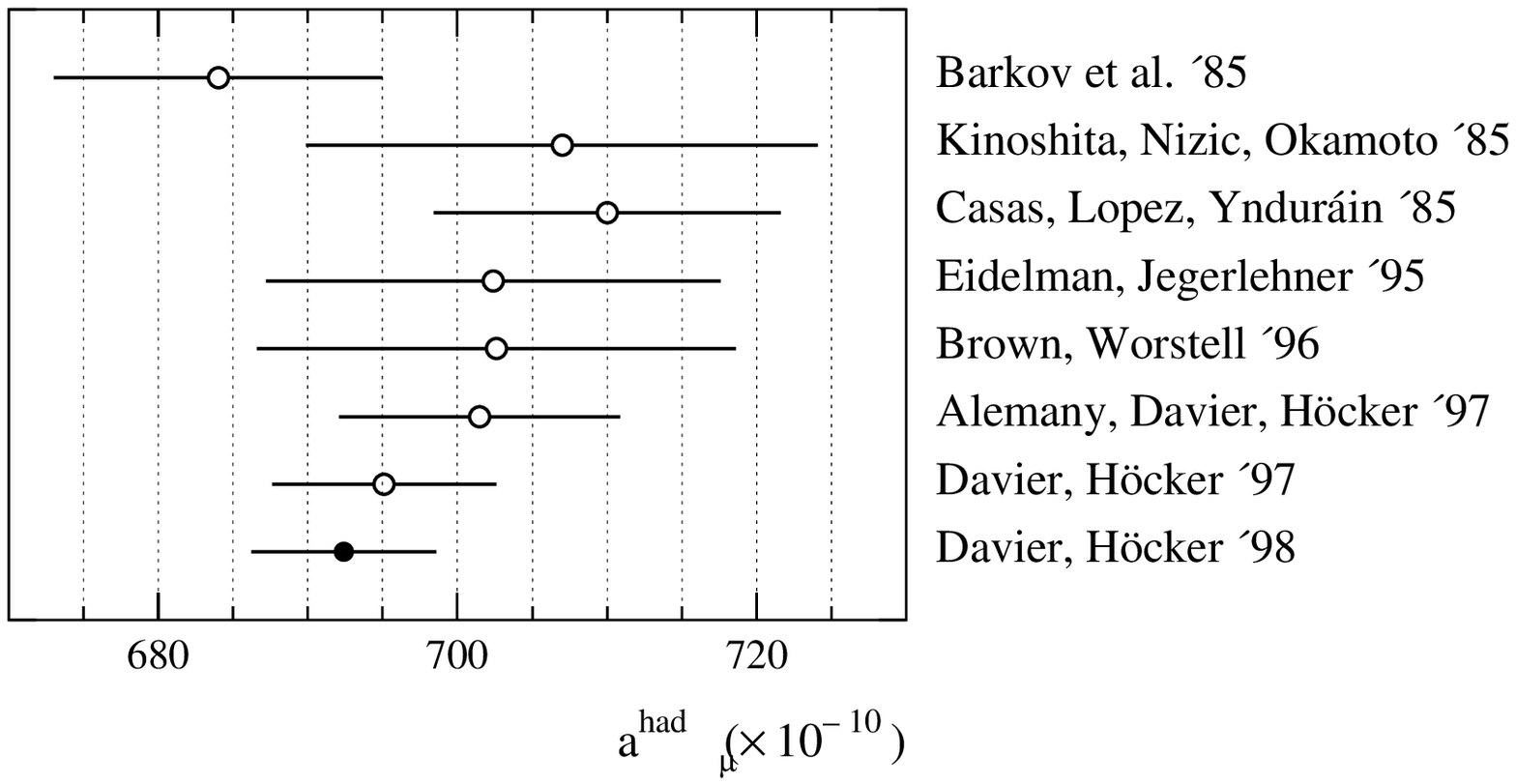,height=1.65in}
\end{tabular*}
\vspace*{-0.5cm}
\caption[.]{\label{fig_results_all}
            Comparison of \daqedhZ\ and $a_\mu^{\rm had}$ evaluations.
            The values for \daqedhZ\ are taken from Refs.~\rm
 \cite{lynn,eidelman,burkhardt,martin,swartz,g_2pap,alphapap,kuhnstein,erler,apap},
            while those for $a_\mu^{\rm had}$ are from Refs.~\rm
            \cite{barkov,kinoshita,casas,eidelman,worstell,g_2pap,alphapap,apap}.}
\end{figure*}

Considering the direct Higgs search currently conducted at CERN
by the four LEP experiments, it is worth remarking that the $21~{\rm GeV/c}^2$
shift between the two indirect determinations is almost entirely
due to the region 2.0 - 3.7 GeV in the evaluation of \daqedh\ where the
available \eee\ data are systematically higher than the QCD prediction
(see Fig.~\ref{fig_r}). In this respect, the preliminary results 
from BES~\cite{ichep_bes}, which are more precise than the earlier 
measurements in this energy range, are observed to nicely agree with QCD.
We are looking forward to more complete and more precise results
in this crucial energy region.

\section{SUMMARIZING THE PROCEDURE AND ITS JUSTIFICATION}

We have described a three-step procedure to improve the evaluation of
hadronic vacuum polarisation occuring in the anomalous magnetic
moments of the leptons and the running of $\alpha$. By far the most
rewarding step was to replace poor experimental data on \eee\ annihilation
cross sections in the 1.8 - 3.7 GeV range and above 5 GeV by a precise
QCD prediction. The justification for believing this prediction at the
level quoted ($\sim 1\%$) follows from direct tests using 
{\it experimental data}.
\newpage
The two assumptions needed for applying QCD to this problem are: {\it (i)}
$R(s)$ can be approximated by $R_{QCD}(s)$ in an 
{\it average} sense only since an integral is computed 
({\it global} quark-hadron duality) and {\it (ii)}
perturbative QCD can be reliably used at energies as low as 1.8 GeV.

These hypotheses have been thoroughly tested in the study of hadronic 
$\tau$ decays~\cite{aleph_asf,opal_asf} for the dominant isovector amplitude, 
integrating from threshold to 1.8 GeV. Using the precise measurement of
$R_{\tau,V+A}$ and moments of the corresponding spectral function, the 
nonperturbative contributions were found to be smaller than $1\%$, thus 
enabling to validate the perturbative QCD prediction from 1.8 down to 1.0 GeV.
Over this range, the precision of the test reaches $\sim 1\%$, going down to
$\sim 2\%$ near 1 GeV. The consistency of the QCD description can be 
expressed through the values of $\alpha_s(M_Z^2)$ found in the different
cases studied: $0.1202 \pm 0.0008_{exp} \pm 0.0024_{th} \pm 0.0010_{evol}$ 
for ($V+A\,,I=1$), $0.1197 \pm 0.0018$ for ($V\,,I=1$), $0.1207 \pm 0.0017$ 
for ($A\,,I=1$) in $\tau$ decays~\cite{aleph_asf}, and $0.1205 \pm 0.0053$ 
for ($V\,,I=0,1$) in \eee\ annihilation~\cite{alphapap}.
 
In essence, the nice properties (quark-hadron duality and validity of
perturbative QCD calculations) observed in an a priori critical energy 
region are applied at higher and safer energies, where the achieved 
precision should be $1\%$ or better.

\section{CONCLUSIONS}

This note summarizes the recent ef\/fort that has been
undertaken in order to ameliorate the theoretical predictions for 
\aqedZ\ and \amuhad, crucially necessary to maintain the 
sensitivity of the diverse experimental improvements on the
Standard Model Higgs mass, on the one hand, and tests of the
electroweak theory on the other hand. The new value of 
$\alpha^{-1}(M_{\rm Z}^2)=128.933\pm0.021$ for the running f\/ine structure 
constant is now suf\/f\/iciently accurate so that its precision is no 
longer a limitation in the global Standard Model f\/it. On the contrary,
more ef\/fort is needed to further improve the precision of the
hadronic contribution to the anomalous magnetic moment of the muon
below the intended experimental accuracy of BNL-E821~\cite{bnl}, 
which is about $4\times10^{-10}$. Fortunately, new low energy 
data are expected in the near future from $\tau$ decays (CLEO, OPAL, 
DELPHI, BaBar) and from \eee\ annihilation (BES, CMD~II, DA$\Phi$NE).
Additional support might come from the theoretical side
using chiral perturbation theory to access the low energy inverse 
moment sum rules (\ref{eq_int_alpha}) and (\ref{eq_int_amu}). One 
could, \eg, apply a similar procedure as the one which was used in 
Ref.~\cite{stern} to determine the constant $L_{10}$ of the chiral 
lagrangian.

\section*{Acknowledgements}
It is indeed a pleasure to thank my collaborator and friend Andreas H\"ocker
for the exciting work we are doing together at LAL. Many interesting 
discussions with J.~K\"uhn are acknowledged . I would like to congratulate
Toni Pich and Alberto Ruiz with their local committee for the perfect 
organisation of this Workshop in Santander.


\begin{thebibliography}{99}
\bibitem{g_2pap}     R.~Alemany, M.~Davier and A.~H\"ocker,
                     \EPJ\ {\bf C2} (1998) 123
\bibitem{eidelman}   S.~Eidelman and F.~Jegerlehner,
                     \ZP\ {\bf C67} (1995) 585
\bibitem{bailey}     J.~Bailey \ea, 
                     \PL\ {\bf B68} (1977) 191. \\
                     F.J.M.~Farley and E.~Picasso, ``{\it 
                     The muon $(g-2)$ Experiments}'',
                     Advanced Series on Directions in High Energy Physics - 
                     Vol.~7 Quantum Electrodynamics, ed. T.~Kinoshita, 
                     World Scientif\/ic 1990
\bibitem{bnl}        B.~Lee~Roberts,
                     \ZP\ {\bf C56} (Proc. Suppl.) (1992) 101
\bibitem{bnlnew}     C.~Timmermans, \ICHEP, Vancouver (1998)
\bibitem{sinth}      D.~Karlen, \ICHEP, Vancouver (1998)
\bibitem{topmass}    R.~Partridge, \ICHEP, Vancouver (1998)
\bibitem{degrassi}   G.~Degrassi, P.~Gambino, M.~Passera and A.~Sirlin,
                     \PL\ {\bf B418} (1998) 209; \\
                     G.~Degrassi, Talk given at the Zeuthen Workshop on 
		     Elementary Particle Theory: Loops and Legs in Gauge 
                     Theories, Rheinsberg, Germany (1998), hep-ph/9807293
\bibitem{steinh}     M.~Steinhauser, 
                     ``{\it 
		       Leptonic contribution to the effective electromagnetic
                       coupling constant up to three loops}'', 
                     Report MPI/PhT/98-22 (1998)
\bibitem{cabibbo}    N.~Cabibbo and R.~Gatto, \PRL\ {\bf 4} (1960) 313;
                     \PR\ {\bf 124} (1961) 1577.
\bibitem{krause1}    A.~Czarnecki, B.~Krause and W.J.~Marciano,
                     \PRL\ {\bf 76} (1995) 3267; \PR\ {\bf D52} (1995) 2619
\bibitem{peris}      S.~Peris, M.~Perrottet and E.~de Rafael, 
                     \PL\ {\bf B355} (1995) 523
\bibitem{weinberg}   R.~Jackiw and S.~Weinberg, \PR\ {\bf D5} (1972) 2473
\bibitem{rafael}     M.~Gourdin and E.~de~Rafael, \NP\ {\bf B10} (1969) 667
\bibitem{rafael2}    S.J.~Brodsky and E.~de Rafael, \PR\ {\bf 168} (1968) 1620
\bibitem{aleph_vsf}  ALEPH \Cl\ (R.~Barate \ea)
                     \ZP\ {\bf C76} (1997) 15
\bibitem{etapi}      S.~Tisserant and T.N.~Truong,
                     {\it Phys. Lett.} {\bf B115} (1982) 264; \\
                     A.~Pich, {\it Phys. Lett.} {\bf B196} (1987) 561; \\
                     H.~Neufeld and H.~Rupertsberger,
                     {\it Z. Phys.} {\bf C68} (1995) 91
\bibitem{pdg}         C.~Caso \ea\ (Particle Data Group),\EPJ\ {\bf C3} (1998) 1
\bibitem{ichep_bes}  Z.~Zhao, \ICHEP, Vancouver (1998)
\bibitem{aleph_asf}  ALEPH \Cl\ (R.~Barate \ea),
                     {\it Eur. Phys. J.} {\bf C4} (1998) 409
\bibitem{opal_asf}   OPAL \Cl\ (K.~Ackerstaf\/f \ea), ``{\it
                     Measurement of the Strong Coupling Constant $\alpha_s$ and the 
                     Vector and Axial-Vector Spectral Functions in
                     Hadronic Tau Decays}'', CERN-EP/98-102 (1998)
\bibitem{wilson}     K.G.~Wilson, \PR\ {\bf 179} (1969) 1499
\bibitem{svz}        M.A.~Shifman, A.L.~Vainshtein and V.I.~Zakharov,
                     \NP\ {\bf B147} (1979) 385, 448, 519
\bibitem{alphapap}   M.~Davier and A.~H\"ocker, \PL\ {\bf B419} (1998) 419
\bibitem{adler}      S.~Adler, \PR\ {\bf D10} (1974) 3714
\bibitem{3loop}      L.R.~Surguladze and M.A.~Samuel, 
                     \PRL\ {\bf 66} (1991) 560; \\
                     S.G.~Gorishny, A.L.~Kataev and S.A.~Larin,
                     \PL\ {\bf B259} (1991) 144
\bibitem{kataev}     S.G.~Gorishny, A.L.~Kataev and S.A.~Larin,
                     {\it Il Nuovo Cim.} {\bf 92A} (1986) 119
\bibitem{bnp}        E.~Braaten, S.~Narison and A.~Pich, 
                     \NP\ {\bf B373} (1992) 581
\bibitem{apap}       M.~Davier and A.~H\"ocker, ``{\it
                     New Results on the Hadronic Contribution to 
                     \aqedZ\ and to $(g-2)_\mu$}'',
                     Report LAL 98-38 (1998), to be published in \PL\
\bibitem{moriond98}  M.~Davier, ``{\it
                     New Results on $\alpha_s(M_Z^2)$ and \aqedZ}'',
                     Report LAL 98-53 (1998), to be published in 
                     the Proceedings of the Rencontres de Moriond
                     ``{\it Electroweak Interactions and Unified Theories}''
                     Les Arcs, France, March 14-21, 1998  
\bibitem{kuhn1}      K.G.~Chetyrkin, J.H.~K\"uhn and M.~Steinhauser,
                     \NP\ {\bf B482} (1996) 213
\bibitem{reinders}   L.J.~Reinders, H.~Rubinstein and S.~Yazaki,
                     \PRep\ {\bf 127} (1985) 1
\bibitem{bertlmann}  R.A.~Bertlmann, \ZP\ {\bf C39} (1988) 231
\bibitem{E_78}       C.~Bacci \ea\ ($\gamma\gamma2$ \Cl),
                     \PL\ {\bf B86} (1979) 234
\bibitem{E_96}       J.L.~Siegrist \ea\ (MARK~I \Cl),
                     \PR\ {\bf D26} (1982) 969
\bibitem{CB}         Z.~Jakubowski \ea\ (Crystal Ball \Cl),
                     \ZP\ {\bf C40} (1988) 49; \\
                     C.~Edwards \ea\ (Crystal Ball \Cl),
                     SLAC-PUB-5160 (1990)
\bibitem{MD1}        A.E.~Blinov \ea\ (MD-1 \Cl),
                     \ZP\ {\bf C49} (1991) 239; \\
                     A.E.~Blinov \ea\ (MD-1 \Cl),
                     \ZP\ {\bf C70} (1996) 31
\bibitem{kuhnstein}  J.H.~K\"uhn and M.~Steinhauser,
                     ``{\it 
 	               A theory driven analysis of the effective QED
                       coupling at $M_{\rm Z}$}'', MPI-PHT-98-12 (1998)
\bibitem{erler}      J.~Erler, ``{\it 
                     \MSbar\ Scheme Calculation of the QED
                     Coupling $\hat{\alpha}(M_Z)$}'', UPR-796-T (1998)
\bibitem{schilcher}  S.~Groote, J.G.~K\"orner, N.F.~Nasrallah and K.~Schilcher,
                     ``{\it 
                       QCD Sum Rule Determination of $\alpha(M_{\rm Z})$
                       with Minimal Data Input}'', 
                     Report MZ-TH-98-02 (1998)
\bibitem{krause2}    B.~Krause, \PL\ {\bf B390} (1997) 392
\bibitem{kinolight}  M.~Hayakawa, T.~Kinoshita, \PR\ {\bf D57} (1998) 465
\bibitem{light2}     J.~Bijnens, E.~Pallante and J.~Prades,
                     \NP\ {\bf B474} (1996) 379
\bibitem{lynn}       B.W.~Lynn, G.~Penso and C.~Verzegnassi,
                     \PR\ {\bf D35} (1987) 42
\bibitem{burkhardt}  H.~Burkhardt and B.~Pietrzyk, \PL\ {\bf B356} (1995) 398
\bibitem{martin}     A.D.~Martin and D.~Zeppenfeld, \PL\ {\bf B345} (1995) 558
\bibitem{swartz}     M.L.~Swartz, \PR\ {\bf D53} (1996) 5268
\bibitem{barkov}     L.M.~Barkov {\it et al.} (OLYA, CMD Collaboration),
                     \NP\ {\bf B256} (1985) 365
\bibitem{kinoshita}  T.~Kinoshita, B.~Ni\v{z}i\'c and Y.~Okamoto,
                     \PR\ {\bf D31} (1985) 2108
\bibitem{casas}      J.A.~Casas, C.~L\'opez and F.J.~Yndur\'ain, 
                     \PR\ {\bf D32} (1985) 736
\bibitem{worstell}   D.H.~Brown and W.A.~Worstell, \PR\ {\bf D54} (1996) 3237
\bibitem{leplib}     D.~Bardin \ea, ``{\it
                     Reports of the working group on precision calculations
                     for the Z resonance''}, CERN-PPE 95-03
\bibitem{stern}      M.~Davier, L.~Girlanda, A.~H\"ocker and J.~Stern,
                     ``{\it 
                        Finite Energy Chiral Sum Rules and Tau Spectral Functions}'',
                     IPNO-TH-98-05, LAL-98-05, hep-ph/9802447, 
                     {\it to be published in Phys. Rev. D}

\end{thebibliography}
\end{document}